\documentclass{article}
\usepackage{spconf,amsmath,epsfig}
\usepackage{multirow}
\usepackage{graphicx}
\usepackage{subfigure}
\usepackage{comment}
\usepackage{color}
\usepackage{amsmath}
\usepackage{bbm}
\usepackage{tikz}
\usepackage{booktabs} 
\usepackage{indentfirst}
\usepackage{multirow}
\usepackage{algorithm}
\usepackage{array}
\usepackage[noend]{algpseudocode}
\algdef{SE}[DOWHILE]{Do}{doWhile}{\algorithmicdo}[1]{\algorithmicwhile\ #1}%
\usepackage{array}
\usepackage{setspace}

\definecolor{orange}{rgb}{1,0.5,0}

\begin{document}\sloppy

\def\x{{\mathbf x}}
\def\L{{\cal L}}

\title{Reactive Video Caching via long-short-term fusion approach}
%
\name{Rui-Xiao Zhang$^{*}$, Tianchi Huang$^{*}$, Xin Yao$^{*}$, Chenglei Wu$^{*}$, Lifeng Sun$^{*}$}

\address{$^{*}$Tsinghua University}

\maketitle

%
\begin{abstract}

Video caching has been a basic network functionality in today's network architectures. Although the abundance of caching replacement algorithms has been proposed recently, these methods all suffer from a key limitation: due to their immature rules, inaccurate feature engineering or unresponsive model update, they cannot strike a balance between the long-term history and short-term sudden events. To address this concern, we propose LA-E2, a long-short-term fusion caching replacement approach, which is based on a learning-aided exploration-exploitation process. Specifically, by effectively combining the deep neural network (DNN) based prediction with the online exploitation-exploration process through a \emph{top-k} method, LA-E2 can both make use of the historical information and adapt to the constantly changing popularity responsively.
Through the extensive experiments in two real-world datasets, we show that LA-E2 can achieve state-of-the-art performance and generalize well. Especially when the cache size is small, our approach can outperform the baselines by 17.5\%-68.7\% higher in total hit rate.

\end{abstract}
\vspace{-5pt}
\section{Introduction}
\label{sec:intro}

Among all content transferred in the network, multimedia, especially video content, has played a dominant part. As reported, video content will account for as much as 75\%~\cite{cisco} of the total network traffic in the future. However, the limited capacity of the backbone network has presented enormous challenges to content providers (CPs). On the one hand, a growing number of video audiences are eager to have more timely and higher quality of viewing experience; on the other hand, CPs need to deliver videos in a more cost-effective way to make more profits. To better serve users with limited bandwidth resources and trim costs, content caching has been accepted as a promising way due to its capacity of traffic offloading.

However, although the caching technique has been used for decades, the constantly changing network conditions have brought new challenges to researchers.
For example, in 5G networks, the caching storage in cellular base stations is becoming increasingly small, and more caching devices will be placed closer to end users~\cite{shanmugam2013femtocaching}. These trends make the content caching problem more heterogeneous. As a result, how to establish a more effective caching policy, especially in small cache size scenarios, is always a hit topic.

Today, most replacement algorithms are rule-based: first in first out (FIFO), least recently used (LRU), least frequently used (LFU) and their variants~\cite{shafiq2014revisiting}.
However, the simplified rules and ignorance of popularity make them hard to adapt to the requests' dynamics~\cite{pang2018toward}. To deal with it, prediction-based algorithms have received more attention in recent years~\cite{li2016popularity}. By making use of historical data, they extract some handcraft features to estimate the popularity of each content and replace the least popular item, i.e., the item with the least populairty. Nevertheless, these algorithms require significant tuning, and the insights found in one scenario may not generalize well in others. To solve these problems, some deep learning methods have been suggested in most recent work~\cite{pang2018toward,narayanan2018deepcache}. By using the deep neural network (DNN), these algorithms can predict popularity without any presumptions. 
Unfortunately, these methods also have a key limitations: due to that the converge of a DNN model needs a lot of data and time as well, its parameters can only be updated periodically on coarse timescales, which means that they cannot detect and adapt to sudden events such as the appearance of new hot contents. 


In this paper, we propose a learning-aided exploration-exploitation (LA-E2) approach to solve caching replacement problem. Specifically, we will highlight two most fundamental challenges: 1) How to design an approach which can well capture the long-term historical information and generalize well to different requests patterns (e.g, popularity distribution). 
2) How to design an approach which can deal with the sudden events such as the appearance of hot contents.

To address these concerns, we argue that the caching replacement problem should be treated as not only a popularity prediction problem, but also an online exploration and exploitation process, that is, using prediction method to extract long-term popularity features from historical information, while using online E2 to deal with short-term popularity changes.
The critical insight behind our approach is that from an empirical observation, we find that even though the DNN model may not adapt to the sudden event appearance and
lead to an accurate prediction of the best replacement, which is the least popular object, it can still make use of historical information to form a small subset of candidates which contains the best choice. LA-E2, then, utilizes an online E2 technique to adaptively identify the most suitable choice in a real-time way. In this way, LA-E2 can strike a balance between the prediction-based approach, which focuses more on long-term popularity information, and exploration-based approach, which focuses more on short-term sudden events. The main contributions are summarized as follows:

1. We propose LA-E2, a caching replacement approach, which can both extract long-term popularity features without any predefined processing and adapt itself to short-term sudden popularity changes. To the best of our knowledge, we are the first to combine the deep learning method with E2 process to solve the caching replacement problem.

2. We analyze the potential of a high-performance replacement approach. We deeply study how to use LA-E2 to alleviate the fundamental problems which previous algorithms still face.


3. Through the extensive experiments in two real-world datasets, we find that LA-E2 can achieve state-of-the-art performance. Especially, when the cache size is small, LA-E2 outperforms the baselines by 8.4\%-72.3\% and 17.5\%-68.7\% in two datasets, respectively.

The remainder of this paper is organized as follows. \S \ref{sec:related} presents related work. Both the system architecture and problem formulation are shown in \S\ref{sec:profile}. The details of LA-E2 are provided in \S \ref{design}. The experiments and conclusions are given in \S \ref{experiment} and \S\ref{conclusion} respectively.\vspace{-6 pt}
\section{related work}\label{sec:related}

Content caching has been a basic network functionality in today's network architectures such as content delivery network (CDN)~\cite{pallis2006insight} and 5G networking~\cite{poularakis2014approximation}. Due to the simplicity of implementation, most currently used caching algorithms are still based on FIFO, LRU, and LFU~\cite{shafiq2014revisiting}. 
However, the effectiveness of these algorithms requires the presumption of the request patterns (e.g., Poisson arrival), which is always problematic in the real world.

At the same time, some data-driven algorithms have been proposed in recent years. Researchers in~\cite{li2016popularity} propose a popularity-based learning algorithm for cache replacement.~\cite{bacstuug2014living} predicts time-varying popularity through collaborative filtering. 
However, since these algorithms need handcraft feature engineering and pre-processing, they cannot be generalized well to deal with different scenarios. 

Inspired by the success of deep learning method, the latest work has paid attention to using DNN to solve the caching problem. In~\cite{pang2018toward}, authors propose an approach named ``DeepCache'', which directly uses long-short-term-memory (LSTM) network to predict popularity. In~\cite{narayanan2018deepcache}, researchers use a sequence-to-sequence model, also based on LSTM, to predict future characteristics of each content. Nevertheless, these algorithms may suffer from prediction bias due to unresponsive model update, which separates them from the optimal solution. \vspace{-5pt}




\begin{figure}[t]
  \centering
  
  \begin{minipage}{1.0\linewidth}
    \centering
      {\includegraphics[width=1.0\textwidth]{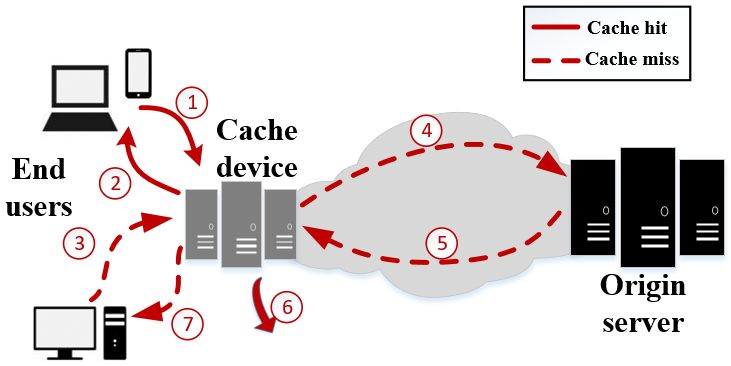}}
  \end{minipage}
  \caption{Video caching system}
  \label{fig:system}
 \end{figure}

\vspace{0pt}
\section{Cache replacement profiles}\label{sec:profile}
\subsection{System overview}
Fig \ref{fig:system} shows a typical caching system. As is shown, the system consists of three parts: the origin server, the cache device, and the end users. The origin server is possessed by a CP and contains all video contents. The cache device is located close to end users and able to be an edge server or a smart router. 
At each time, when an end user requests a video content (\textcircled{1}/\textcircled{3}), it will first go to the cache device and see whether the content can be accessed. If accessible (\textcircled{2}), the request will be directly served, which is called a cache hit; otherwise, a cache miss occurs, and the cache device will go to the origin server (\textcircled{4}) to fetch the content (\textcircled{5}), and then a replacement is needed to make room for it (\textcircled{6}). Finally, the user will get served by the new content(\textcircled{7}).
\vspace{-10pt}

\subsection{Problem formulation}
Without loss of generality, we only consider one single cache device, which can be easily extended to multiple ones. Suppose that a CP has an $N$-contents set, which is denoted as $\boldsymbol{C}=\{1, 2, ...i,..., N\}$, and $\boldsymbol{R} = \{R_t, t\in \mathbf{Z} \}$ represents the successive contents requested by end users. Just like~\cite{li2016popularity,pang2018toward}, in this paper we also regard all contents as the same size, and the cache device can only cache up to $K$ different contents. In practice, $K$ is much less than $N$. The cached contents at timeslot $t$ are denoted as $\boldsymbol{\hat{C(t)}}=\{\hat{C_{1}(t)},\hat{C_{2}(t)},...,\hat{C_{K}(t)} \}$.

At the same time, for each request $R_t$, we use an indicator $\mathbbm{1}$ to identify whether it is a cache hit or a cache miss, which satisfies $\mathbbm{1} \in \{0,1\}$. When it is a cache hit denoted as $R_t \in \boldsymbol{\hat{C(t)}}$, we set $\mathbbm{1}=1$. Otherwise $\mathbbm{1}=0$, and a replacement is needed. Here we refer the evicted content as $e_t$, which is determined by the replacement algorithm. The objective of the problem is to get as many cache hits as possible. Formally, the caching problem is defined as follows:
\begin{align}
&\max \limits_{e_t}  \frac{1}{T} \sum_{t=1}^T {\mathbbm{1}_{\{R_t \in \boldsymbol{\hat{C(t)}}\}}}\\
&s.t.\quad
\begin{cases}
K \ll N, \\
e_t \in \{\hat{C_{1}(t)},\hat{C_{2}(t)},...,\hat{C_{K}(t)}\}.
\end{cases}
\end{align}

\begin{figure}
	\centering
    \begin{minipage}{0.48\linewidth}
    	\centering
  		\subfigure[Comparison of existing algorithms in two real-world dataset]{\includegraphics[width=1.0\linewidth]{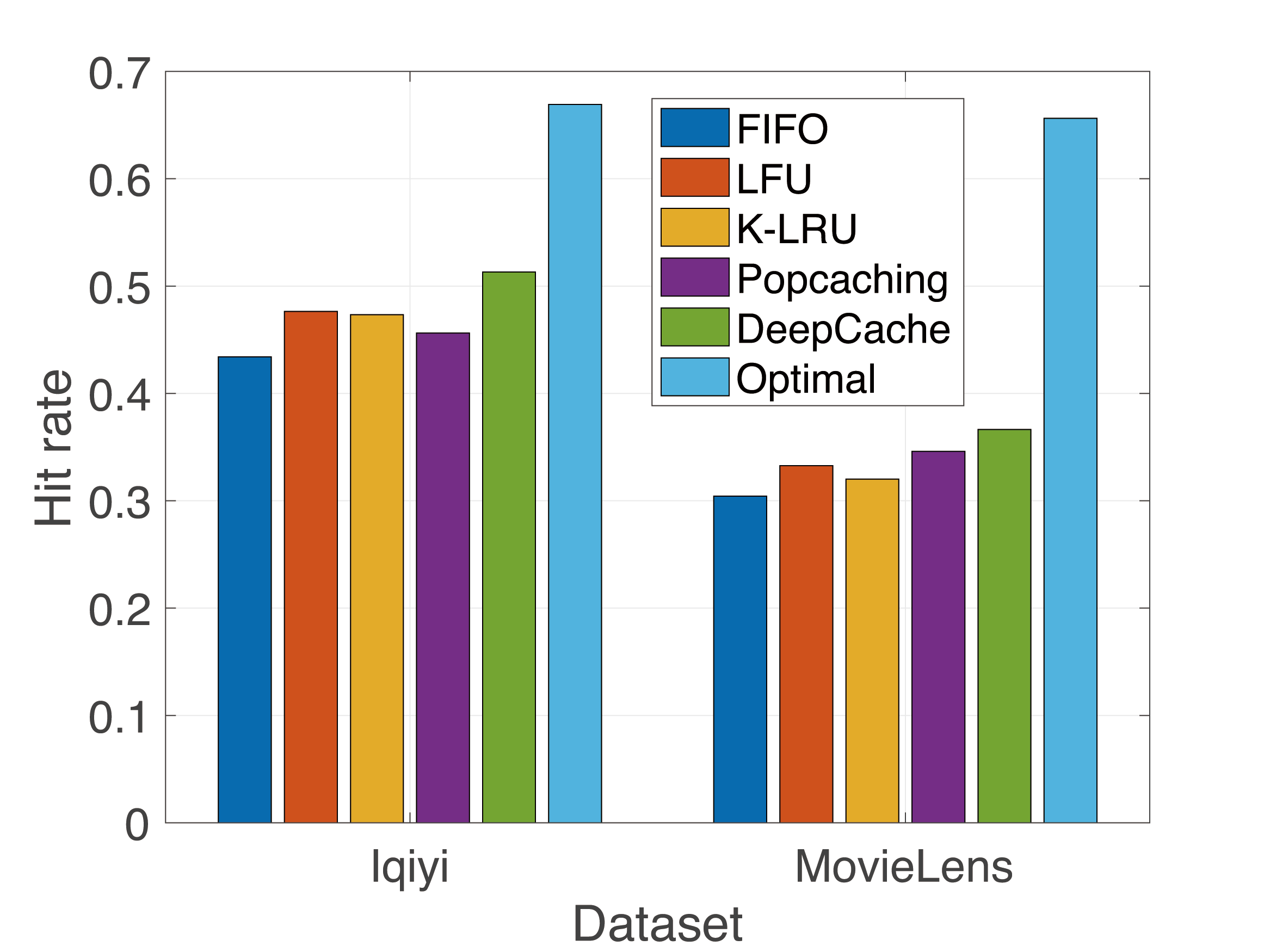}}
     \end{minipage}
    \begin{minipage}{0.48\linewidth}
    	\centering
  		\subfigure[The distribution of \emph{the optimal solution is in top-k prediction results} ]{\includegraphics[width=1.0\linewidth]{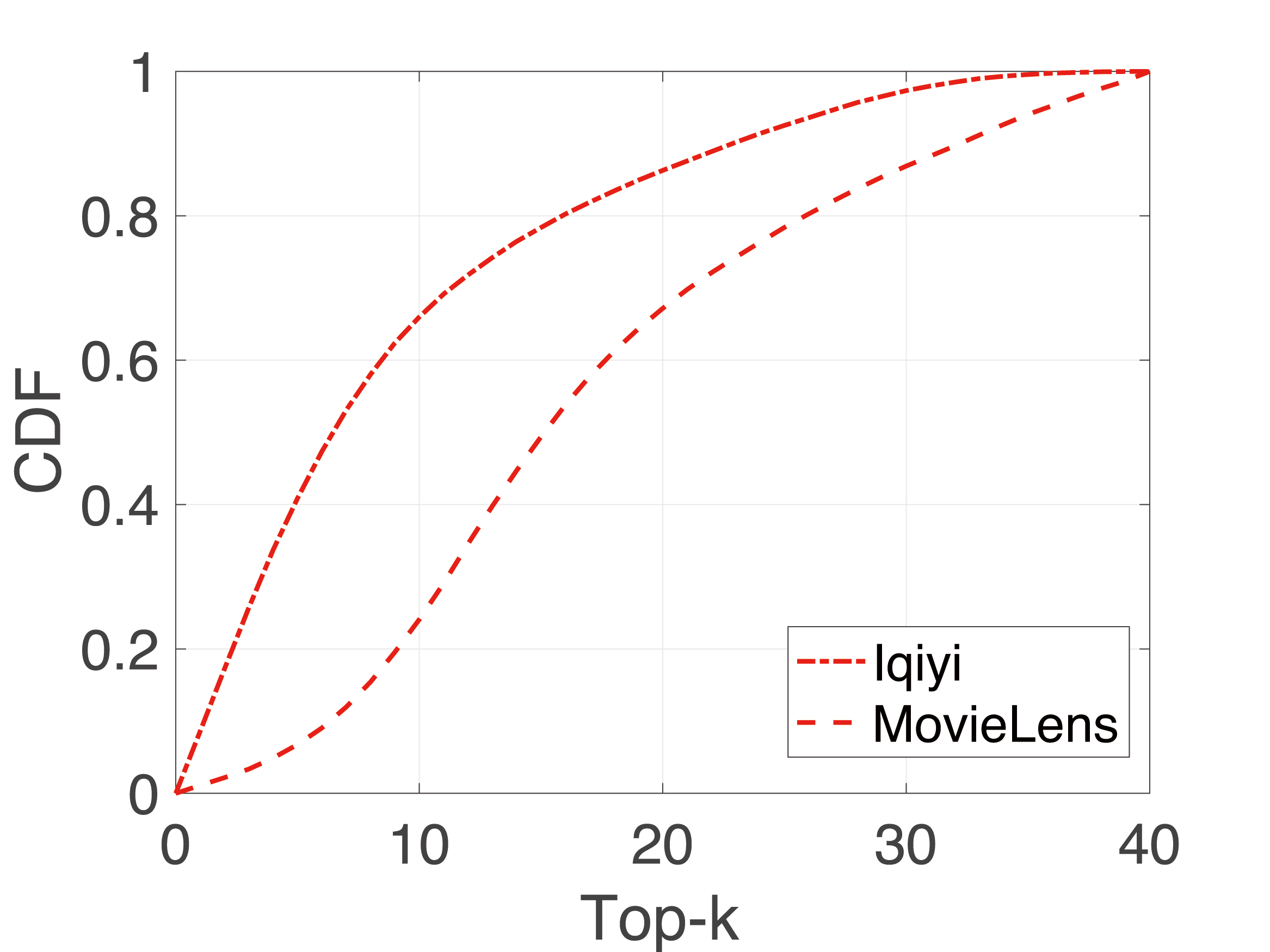}}
     \end{minipage}     
   
   \caption{The potential of a better replacement algorithm}
    \label{fig:motivation} 
\end{figure}

\vspace{-5pt}
\subsection{The potential of a better replacement approach}\label{sec:potential}
To design a better replacement approach, we first compare existing algorithms in two real-world datasets (for the completeness of paper structure, we present the profiles of the datasests and algorithms in \S \ref{experiment:dataset} and \S \ref{experiment:baselines}, respectively). Notably, by using the optimal solution~\cite{mattson1970evaluation}, which needs to know all the future requests sequence and cannot be obtained in reality, we quantify the potential gain of a better approach. 

The algorithms involved in the comparison include three types: the rule-based (FIFO, K-LRU, LFU), the traditional prediction-based (Popcaching)~\cite{li2016popularity}, and the DNN-based (DeepCache)~\cite{pang2018toward}. Their performance is depicted in Fig \ref{fig:motivation}(a), and we can find the following observations:
\begin{itemize}
    \item Comparing with traditional rule-based and prediction-based algorithms, DeepCache performs the best in both Iqiyi dataset and MovieLens dataset. 
    \item The state-of-the-art algorithm can only reach 40\%-70\% performance of the optimal in hit ratio, and there is a large space for improvement. 
\end{itemize}
\indent The outstanding performance of DeepCache in first observation can be explained:
the rule-based algorithms ignore or only consider the current popularity of the content that actually cannot be directly treated as the future popularity. For instance, FIFO doesn't consider the future popularity, while LFU only considers the popularity in a specific time window. At the same time, we can find that comparing with LFU, Popcaching performs better in MovieLens dataset but worse in Iqiyi dataset. It means that although Popcaching uses feature engineering to predict popularity, the features are still
handcraft and cannot generalize well to deal with different requests patterns. Instead, DNN-based DeepCache can make decisions almost only relying on raw input data, so it can generalize to different datasets. 

At the same time, to explain the second observation, we dive deeper into the DNN-based DeepCache, the best performer. 
We present the distribution of \emph{the optimal solution is in top-k prediction results} in Fig \ref{fig:motivation}(b). We find that the DNN-based algorithm cannot narrow down to the single optimal replacement option in neither Iqiyi dataset nor MovieLens dataset. 
However, the optimal one is often among the top-k predicted options. As is shown in Fig \ref{fig:motivation}(b), in Iqiyi dataset, the probability of the optimal replacement included in top-8 is more than 60\%, 
while 5\% if we only pick the option with the minimal popularity (top-1). Similar observations can also be obtained in MovieLens dataset. 

The reason behind the prediction bias of \emph{top-1} is that
the converge of a DNN model needs both a large amount of data and time as well, so the model should be updated periodically on coarse timescales. However, the request popularity is constantly changing, which means the model updated in coarse time cannot identify the sudden events such as the appearance of new hot contents. Meanwhile, as the long-term popularity can be captured regardless of the update time, the DNN-based algorithm can still predict the \emph{top-k} well based on historical information.

Inspired by the above observations, we conclude that instead of only using a DNN-based predictor, which may suffer from imprecise prediction due to its coarse time update, we should also use an online exploration-exploitation (E2) process to trim the prediction bias and adapt to the real-time popularity changes.

\vspace{-5pt}

\section{The design of LA-E2}\label{design}
\subsection{Overview of LA-E2}

\begin{figure}[t]
  \centering
  
  \begin{minipage}{1.0\linewidth}
    \centering
      {\includegraphics[width=1.0\textwidth]{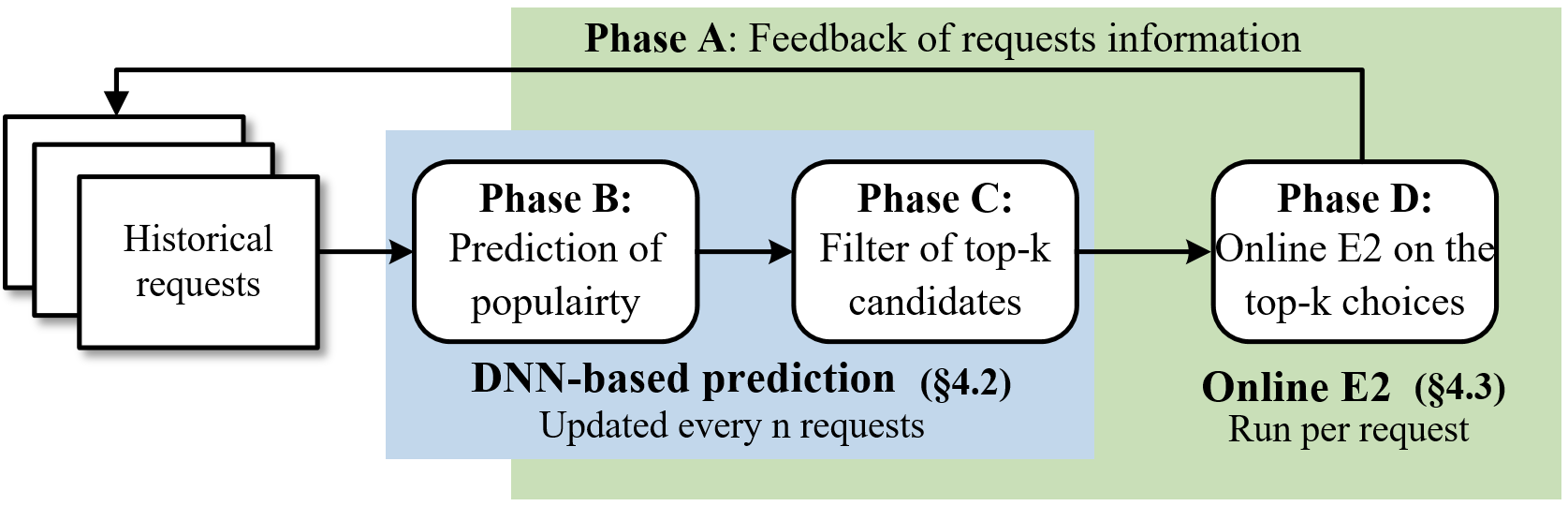}}
  \end{minipage}
  \caption{Overview of LA-E2 caching replacement approach}
  \label{fig:pipline}
 \end{figure}

The intuition behind combining DNN-based prediction and online E2 technique is that an only prediction-based approach cannot attend to the sudden events responsively as it needs a long time to update, while an only online E2-based approach cannot well capture the long-term popularity patterns as it estimates the popularity in a very simple way (e.g., average). 
Following this idea, we pick the \emph{top-k} unpopular alternatives from the DNN-based predictor, which contain the long-term popularity information. Then, we employ the online E2 process to identify the least popular one that combines the short-term popularity information. As a result, LA-E2 is enabled to both consider long-term historical request patterns and adapt itself to short-term popularity changes.
Fig \ref{fig:pipline} presents the main phases in our approach, and the pipeline can be summarized as follows:
\begin{itemize}

\item \textbf{Phase A:} Collecting historical request information.
\item \textbf{Phase B:} Inputting historical information to the DNN to predict the future popularity of each content.
\item \textbf{Phase C:} Selecting the top-k candidates through the prediction results.
\item \textbf{Phase D:} online E2 on the top-k replacement options using multi-armed bandit (MAB) techniques.
\end{itemize}
To this end, the recent requests will be stored as the historical information and fed back in \textbf{Phase A}. \textbf{Phase B, C} are enforced through DNN-based prediction (shown in blue), which are updated on coarse timescales (every $n$ requests), and \textbf{Phase D} is enforced by online E2 process (shown in grey), which is run and updated per request.
We won't describe \textbf{Phase A} due to its clarity. 
The details of \textbf{Phase B, C} are shown in \S \ref{sec:DNN-based}, and we discuss \textbf{Phase D} later in \S \ref{onlineE2}. 

\vspace{-7 pt}
\subsection{DNN-based prediction}\label{sec:DNN-based}
\textbf{Phase B: Popularity prediction }
In this phase, LA-E2 will input the historical information collected in \textbf{Phase A} and output the popularity prediction of each content. 
Inspired by the recent success of the deep learning method in sequence prediction area~\cite{pang2018toward,narayanan2018deepcache}, we use LSTM, a widely used RNN structure, as the predictor.
By using built-in forget and memory gate, LSTM can extract popularity features directly from raw input data, and generalize to different request patterns by itself as well.  

We apply \emph{softmax} function to represent the popularity distribution of each video content~\cite{pang2018toward}. We use cross entropy to estimate the loss and train our neural network. For space limitations, we recommend readers to~\cite{sundermeyer2012lstm,pang2018toward} for more LSTM's training details. As a default setting, we use a 3-layer LSTM, each with 128 hiddent units, and the model is updated every 1000 requests.

\textbf{Phase C: Top-k candidates filtering }
Whenever a replacement is needed, we will evict the least popular video content. Instead of using the single replacement option in \textbf{Phase B}, LA-E2 will sort the cached items by predicted popularity in ascending order and pick the \emph{top-k} unpopular candidates. Through this phase, LA-E2 can select out the candidates based-on long-term historical information.

\begin{figure*}[t]
  \centering
  \begin{minipage}{0.32\linewidth}
      \centering
      \subfigure[Performance comparison in Iqiyi dataset]{\includegraphics[width=1.0\textwidth]{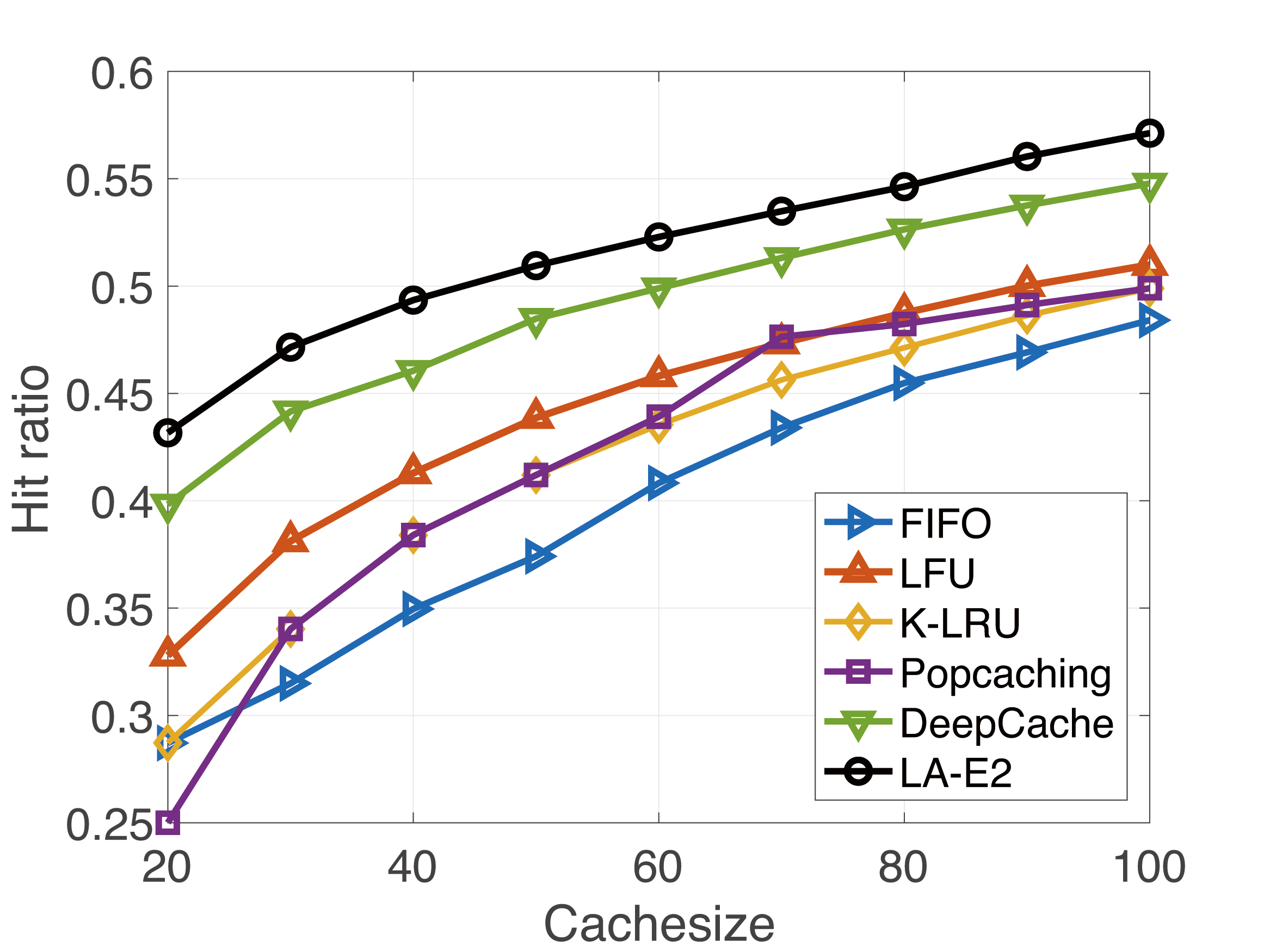}}
  \end{minipage}
  \begin{minipage}{0.32\linewidth}
      \centering
      \subfigure[Performance comparison in MovieLens dataset]{\includegraphics[width=1.0\textwidth]{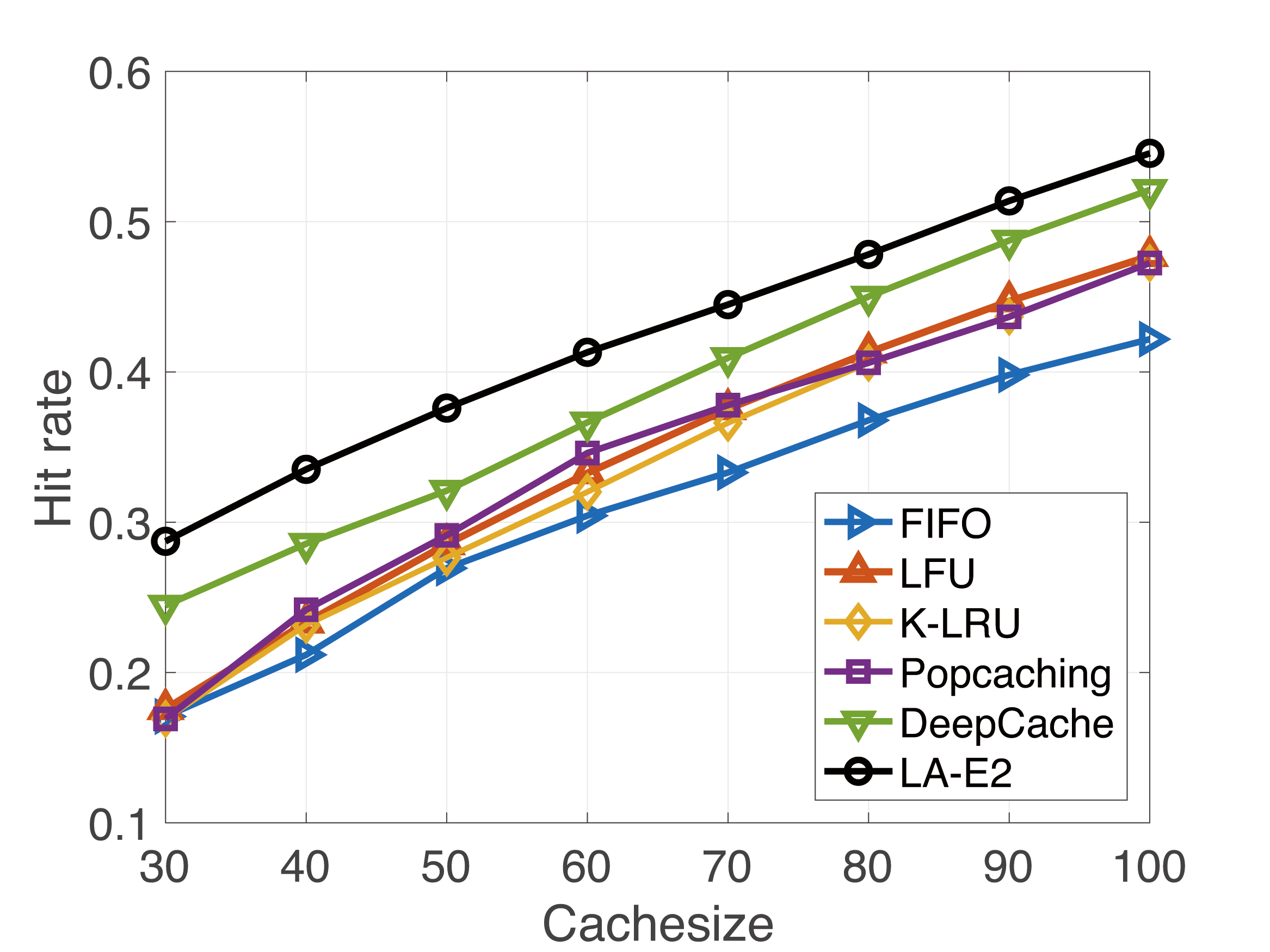}}
  \end{minipage}  
   \begin{minipage}{0.32\linewidth}
     \centering
     \subfigure[Self-comparison experiment]{\includegraphics[width=1.0\textwidth]{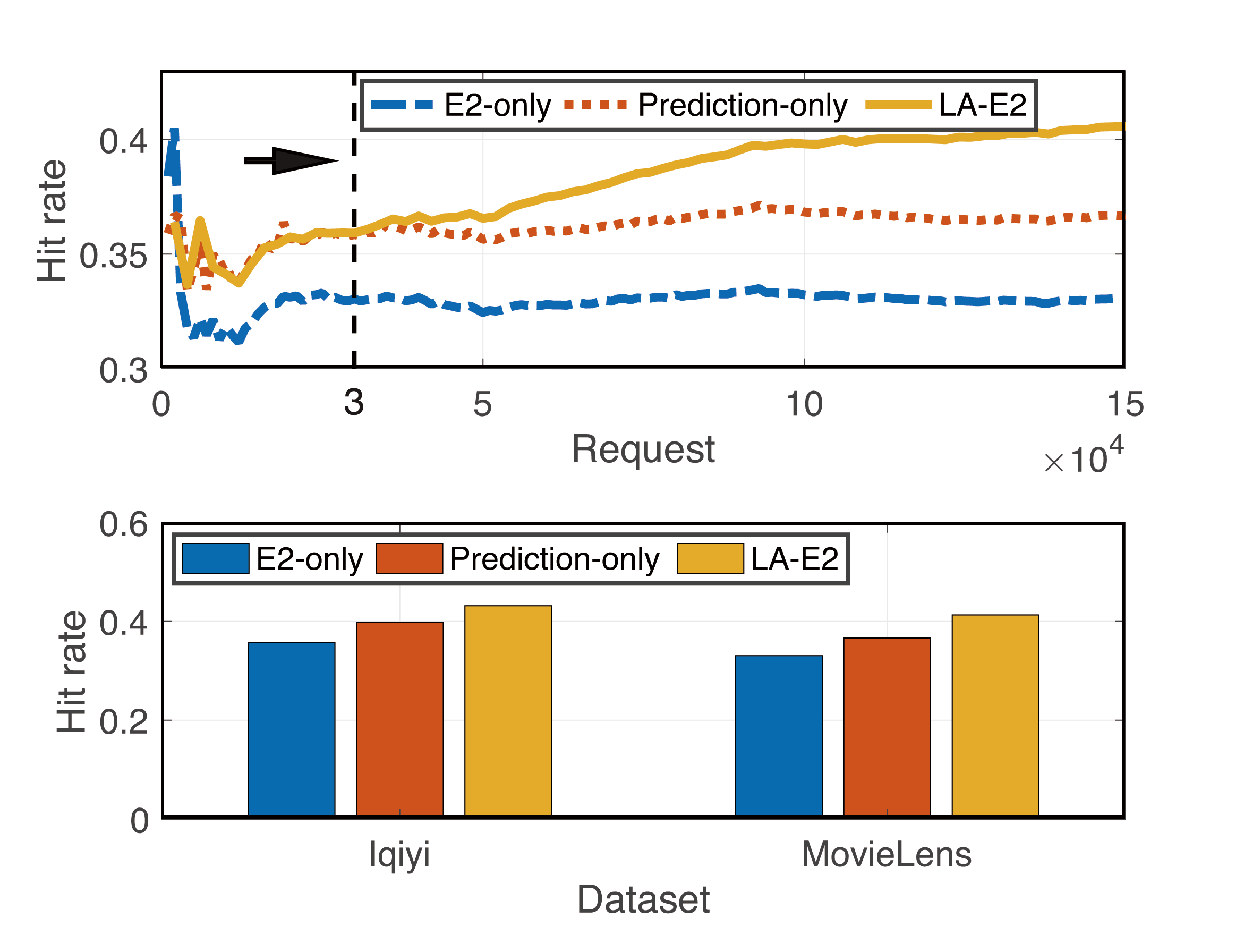}}
  \end{minipage} 
  \caption{LA-E2 can achieve state-of-the-art performance in both datasets. It outperforms the baselines by 8.3\%-72.3\% and 17.5\%-68.7\% higher in two datasets respectively, especially when cache size is small. At the same time, the self-comparison experiment illustrates that both prediction and online E2 are essential.}
  \label{fig:comparison}
\end{figure*}

\subsection{Phase D: Online E2 }\label{onlineE2}
\textbf{Multi-armed bandit problem:} We first make a brief introduction to the Multi-armed-bandit (MAB) problem. 
Consider a bandit with a certain number of arms, and for each arm, the reward distribution is unknown. At each timeslot, one gambler needs to decide which arm should be pulled, and the object is to optimize the long-term average reward. To handle this problem, UCB (Upper Confidence Bound) algorithms are proposed. 
The key idea behind UCB algorithms is to always opportunistically choose the arm with highest upper confidence bound of reward, which will naturally tend to use arms with high expected rewards (exploitation) or with high uncertainty (exploration).


\textbf{UCB for caching replacement: }We first illustrate that the caching replacement problem is a generalization of the classical MAB problem. Each content in cache corresponds to an arm. In other words, there are $K$ arms in total. At each time, LA-E2 observes the popularity of each content, which is equivalent to the reward, and makes a decision of replacement, which is equivalent to pulling an arm. Since the popularity of different contents varies with time, in this paper we use sliding-window UCB (SW-UCB) algorithm~\cite{garivier2008upper} to adapt to this non-stationary process. At timeslot $t$, for each content $i$, the policy constructs a $UCB$, which contains two parts as follows:
\begin{align}
    UCB_{t,i}=\bar{X_t}(\gamma, i)+P_t(\gamma,i)
\end{align}
\noindent in which the first part is the average empirical popularity of content $i$ in a fixed-size horizon $\tau$, and the second part is the padding function~\cite{garivier2008upper} which is used to represent the uncertainty of the popularity estimation. The average empirical popularity is defined as:
\begin{align}
    \bar{X_t}(\tau, i)=\frac{1}{\tau}\sum\limits_{s=t-\tau+1}^{t}\gamma^{t-s}\mathbbm{1}_{\{R_s=i\}}
\end{align}
\noindent here, $\gamma \in (0,1)$ is a discounted factor. The use of $\gamma$ is consistent with the fact that the content that arrives more recently is more likely to be popular in future. The padding function is defined as
\begin{align}
  P_t(\tau,i)=B\sqrt{\frac{\log(t\wedge \tau)}{N_t(\tau,i)}}
\end{align}
\noindent where $t\wedge\tau$ denotes the minimum of $t$ and $\tau$, and $B$ is a hyper parameter. 
$N_t(\tau,i)$ is the number of times that content $i$ has been replaced in recent horizon $\tau$, and the smaller it is, the longer content $i$ has stayed in cache, and it can be calculated as:
\begin{align}
  N_t(\tau,i)=\sum\limits_{s=t-\tau+1}^{t}\mathbbm{1}_{\{i= e_s\}}
\end{align}
It is notable that different from classical MAB problem aiming to choose the arm which has the maximal expected reward, we aim to choose (i.e., evict) the arm (i.e., content) which has the minimum expected reward (i.e., popularity), so the parameter B should be negative, and the policy in \textbf{Phase D} can be formulated as:
\begin{align}
    e_t = \arg \min\limits_{i}\bar{X_t}(\gamma, i)+B\sqrt{\frac{\log(t\wedge \tau)}{N_t(\tau,i)}}, B< 0
\end{align}
From the formulation, we can see that the content with low popularity (i.e., small $\bar{X_t}(\gamma, i)$)and the content that has stayed in cache for a long time (i.e., small $N_t(\tau,i)$), are likely to be replaced. The former is the exploitation process, while the latter is the exploration process. 
This process can be clearly interpreted: if a content suddenly becomes popular, it will be less likely to be replaced, as it has been accessed for many times and is attached with a higher $\bar{X_t}$. Meanwhile, if a content has been replaced for many times (i.e., a large $N_t(\tau,i)$), it should have been treated as a popular content. Since only the cached content can be replaced,
it has been at least requested for $N_t(\tau,i)-1$ times. Therefore, a large $N_t(\tau,i)$ should also be less likely to be replaced, and LA-E2 will tend to replace other contents that hasn't been replaced, which is an exploration. Due to the E2 process is run per request, LA-E2 can deal with the sudden popularity changes responsively. 

\vspace{-10 pt}
\section{Experiment}\label{experiment}
\vspace{-10 pt}
\subsection{Dataset} \label{experiment:dataset}

\vspace{-2 pt}
\textbf{Iqiyi dataset:} The dataset is collected from Iqiyi\footnote{www.iqiyi.com}, which is one of the most popular CPs in China. The time span of the dataset is 2 weeks, and it contains more than 1 billion requests from over 2 million viewers, and the unique video contents are more than 0.2 million.


\textbf{MovieLens dataset:} MovieLens\footnote{movielens.org} is a website in which users can rate and make comments on different movies. Inspired by the previous work~\cite{li2016popularity}, we also use video comments data at different times to simulate requests sequence. 
\vspace{-7 pt}
\subsection{Baseline algorithms} \label{experiment:baselines}

\textbf{FIFO}: The first-in-first-out algorithm which is rule-based. The algorithm will record the request time of each content. When the replacement is needed, the earliest cached item will be evicted first.

\textbf{K-LRU}: A variation of least recently used algorithm which is rule-based. The algorithm is that when the cache device has been already full of data, the data that have not been requested for the longest time will be evicted first. Its modification K-LRU is to solve the ``cache pollution problem''. The core idea of K-LRU is to modify the criterion of ``recently used once'' to ``recently used K times''.

\textbf{LFU}: The least-frequently-used algorithm which is rule-based. The algorithm will keep track of the requested content, and by calculating the access frequency in a certain time window, the content which is requested for the least times will be evicted first. 

\textbf{Popcaching}~\cite{li2016popularity}: The prediction-based algorithm. This algorithm first makes feature extraction through historical requests, through which it learns the relationship between the future popularity of a content and its recent requests patterns.

\textbf{DeepCache}~\cite{pang2018toward}: The state-of-the-art algorithm, which is DNN-based. This algorithm predicts the popularity of each content by using deep LSTM network. After inputting the recent requests sequence, DeepCache will evict the least popular content.



\begin{table}
    \centering
    \begin{tabular}{|c|c|c|c|c|c|}
        \hline
         \textbf{\emph{k}} & 5 & 7 & 10 & 15 & 20   \\
         \hline
         \textbf{Iqiyi} & $0.413$ & $0.424$ & $\textbf{0.432}$ & $0.381$ & $0.357$ \\
         \hline
          \textbf{MovieLens} & $0.387$ & $0.395$ & $0.402$ & $\textbf{0.413}$ & $0.381$ \\
         \hline
    \end{tabular}
    \caption{Cache hit rate under different \emph{top-k}}
    \label{tab:my_label}
\end{table}
\vspace{-7 pt}
\subsection{Evaluation and discussion}
We first discuss the influence of parameter \emph{k} of \emph{top-k} in Table \ref{tab:my_label}. A larger \emph{k} means LA-E2 will make decision depending more on online E2 process, thus focusing more on short-term popularity information, while a smaller \emph{k} will enforce it to pay more attention to long-term popularity. Thus there is a trade-off between long-term prediction and short-term adaptation. In our experiment, we choose $\emph{k}=10$ and $\emph{k}=15$, which is the best choice for two datasets respectively. Since MovieLens is from users' comments, it is more likely to be influenced by recent topics, thus it needs a larger \emph{k} to focus more on short-term popularity changes.



Fig \ref{fig:comparison}(a) and Fig \ref{fig:comparison}(b) show the overall hit rate of LA-E2 and other baseline algorithms under various cache sizes. We can see that our proposed LA-E2 outperforms all the benchmarks in both Iqiyi dataset and MovieLens dataset, thus generalizing well. The performance improvement of LA-E2 can achieve 8.3\%-17.5\% against the second best algorithm and 50.32\%-72.3\% against others, especially when the cache size is small.

To better illustrate LA-E2 approach, we also make a self-comparison experiment. In the upper part of Fig \ref{fig:comparison}(c), we show an example of the overall cache hit rate versus the coming requests in three self comparison groups: E2-only group, prediction-only, and LA-E2 group. Before the 30000-th request comes, LA-E2 and prediction-only group just apply the prediction phase of LA-E2, and as expected, their performance is almost the same. Then, both prediction and online E2 are applied to LA-E2 group, and as depicted in Fig \ref{fig:comparison}(c), the LA-E2 group starts to increase significantly and continuously outperforms other groups. Another observation is that E2-only group significantly outperforms the other two groups at the beginning of the experiment and then starts to suffer from performance degradation. This can be explained: at the beginning, there are no enough data for the DNN predictor to converge, and thus the two groups get an unsatisfactory result. After collecting enough historical requests, the strength of popularity extraction of DNN becomes obvious, and the DNN starts to work, and at the same time, E2-only group keeps in a low performance as it cannot well extract popularity features from long-term historical information. 

In the lower part of \ref{fig:comparison}(c), we show the performance of three self-comparison groups in two datasets. As depicted, in Iqiyi dataset, E2-only group and prediction-only group decrease the total hit rate by $20.92\%$ and $8.38\%$, respectively, while for MovieLens dataset, the decrease comes to $24.96\%$ and $12.72\%$. This observation illustrates that both prediction and online E2 are essential to LA-E2. 

\vspace{-10pt}
\section{Conclusion}\label{conclusion}\vspace{-5pt}
In this paper, we propose a novel approach called LA-E2 to solve the caching replacement problem. By using a DNN predictor, LA-E2 can extract long-term history features and predict popularity without pre-determined processing. At the same time, by using online E2 process, LA-E2 can adapt itself to the constantly changing populairty. The extensive simulation on two real-world datasets proves that LA-E2 can significantly outperform state-of-the art algorithms, which illustrates the effectiveness and generalization as well.
\vspace{-5 pt}

\setstretch{0.87}
\bibliographystyle{IEEEbib}

\bibliography{main.blb}

\begin{thebibliography}{10}

\bibitem{cisco}
Cisco Visual~Networking Index,
\newblock ``Cisco visual networking index: Global mobile data traffic forecast
  update, 2016–2021 white paper,'' 2016.

\bibitem{shanmugam2013femtocaching}
Karthikeyan Shanmugam, Negin Golrezaei, Alexandros~G Dimakis, Andreas~F
  Molisch, and Giuseppe Caire,
\newblock ``Femtocaching: Wireless content delivery through distributed caching
  helpers,''
\newblock {\em IEEE Transactions on Information Theory}, vol. 59, no. 12, pp.
  8402--8413, 2013.

\bibitem{shafiq2014revisiting}
Muhammad~Zubair Shafiq, Alex~X Liu, and Amir~R Khakpour,
\newblock ``Revisiting caching in content delivery networks,''
\newblock in {\em ACM SIGMETRICS Performance Evaluation Review}. ACM, 2014,
  vol.~42, pp. 567--568.

\bibitem{pang2018toward}
Haitian Pang, Jiangchuan Liu, Xiaoyi Fan, and Lifeng Sun,
\newblock ``Toward smart and cooperative edge caching for 5g networks: A deep
  learning based approach,''
\newblock in {\em Proc. of IEEE/ACM International Symposium on Quality of
  Service (IWQoS)}, 2018.

\bibitem{li2016popularity}
Suoheng Li, Jie Xu, Mihaela Van Der~Schaar, and Weiping Li,
\newblock ``Popularity-driven content caching,''
\newblock in {\em Computer Communications, IEEE INFOCOM 2016-The 35th Annual
  IEEE International Conference on}. IEEE, 2016, pp. 1--9.

\bibitem{narayanan2018deepcache}
Arvind Narayanan, Saurabh Verma, Eman Ramadan, Pariya Babaie, and Zhi-Li Zhang,
\newblock ``Deepcache: A deep learning based framework for content caching,''
\newblock in {\em Proceedings of the 2018 Workshop on Network Meets AI \& ML}.
  ACM, 2018, pp. 48--53.

\bibitem{pallis2006insight}
George Pallis and Athena Vakali,
\newblock ``Insight and perspectives for content delivery networks,''
\newblock {\em Communications of the ACM}, vol. 49, no. 1, pp. 101--106, 2006.

\bibitem{poularakis2014approximation}
Konstantinos Poularakis, George Iosifidis, Leandros Tassiulas, et~al.,
\newblock ``Approximation algorithms for mobile data caching in small cell
  networks.,''
\newblock {\em IEEE Trans. Communications}, vol. 62, no. 10, pp. 3665--3677,
  2014.

\bibitem{bacstuug2014living}
Ejder Ba{\c{s}}tu{\u{g}}, Mehdi Bennis, and M{\'e}rouane Debbah,
\newblock ``Living on the edge: The role of proactive caching in 5g wireless
  networks,''
\newblock {\em arXiv preprint arXiv:1405.5974}, 2014.

\bibitem{mattson1970evaluation}
Richard~L Mattson, Jan Gecsei, Donald~R Slutz, and Irving~L Traiger,
\newblock ``Evaluation techniques for storage hierarchies,''
\newblock {\em IBM Systems journal}, vol. 9, no. 2, pp. 78--117, 1970.

\bibitem{sundermeyer2012lstm}
Martin Sundermeyer, Ralf Schl{\"u}ter, and Hermann Ney,
\newblock ``Lstm neural networks for language modeling,''
\newblock in {\em Thirteenth annual conference of the international speech
  communication association}, 2012.

\bibitem{garivier2008upper}
Aur{\'e}lien Garivier and Eric Moulines,
\newblock ``On upper-confidence bound policies for non-stationary bandit
  problems,''
\newblock {\em arXiv preprint arXiv:0805.3415}, 2008.

\end{thebibliography}


\end{document}